\documentclass[12pt]{iopart}

\usepackage{iopams,graphicx,color}
\begin{document}

\title{Multimode entanglement in coupled cavity arrays}

\author{T C H Liew$^{1,2}$ and V Savona$^2$}

\address{$^1$ School of Physical and Mathematical Sciences, Nanyang Technological University, 637371, Singapore}
\address{$^2$ Institute of Theoretical Physics, Ecole Polytechnique F\'{e}d\'{e}rale de Lausanne EPFL, CH-1015 Lausanne, Switzerland}

\ead{tchliew@gmail.com, vincenzo.savona@epfl.ch}

\begin{abstract}
We study a driven-dissipative array of coupled nonlinear optical resonators by numerically solving the Von Neumann equation for the density matrix. We demonstrate that quantum correlated states of many photons can be generated also in the limit where the nonlinearity is much smaller than the losses, contrarily to common expectations. Quantum correlations in this case arise from interference between different pathways that the system can follow in the Hilbert space to reach its steady state under the effect of coherent driving fields. We characterize in particular two systems: a linear chain of three coupled cavities and an array of eight coupled cavities. We demonstrate the existence of a parameter range where the system emits photons with continuous-variable bipartite and quadripartite entanglement, in the case of the first and the second system respectively. This entanglement is shown to survive realistic rates of pure dephasing and opens a new perspective for  the realization of quantum simulators or entangled photon sources without the challenging requirement of strong optical nonlinearities.
\end{abstract}

\pacs{42.50.-p, 71.36.+c, 42.50.Ex}

%42.50.-p: Quantum Optics
%71.36.+c: Polaritons
%42.50.Ex: 	Optical implementations of quantum information processing and transfer

\maketitle

\section{Introduction} The physics of interacting Bose particles on a lattice offers a general framework within which several systems in condensed matter, atomic and optical physics are modelled. In the field of ultracold atoms in particular, the striking experimental demonstrations of phases with strong quantum correlations -- such as the Mott insulator phase \cite{Greiner2002,Bakr2010,Sherson2010} -- and of the associated phase transitions, have convincingly opened the way to the physical realization of quantum simulators \cite{Lloyd1996,Bloch2012}. A similar perspective \cite{Aspuru2012} has vigorously stimulated the fields of quantum optics and cavity quantum electrodynamics (CQED), supported by the technological advances in the fabrication of optical micro- and nano-resonators with exceptionally long photon lifetimes. In this context, a major objective consists in the implementation of the physics of strongly correlated bosons in a system of photons confined in an array of coupled optical resonators. In such a system, the effective on-site interaction between photons can originate from optical nonlinearities, due to some nonlinear medium embedded in each cavity. In particular, two typical kinds of nonlinearities are almost always considered. The first is the nonlinearity at the heart of the Jaynes-Cummings model, arising when each cavity embeds a two-level system, and gives rise to the model known as the {\em Jaynes-Cummings-Hubbard (JCH)} model \cite{Makin2008}. The second is the standard third order optical nonlinearity. It can arise from the nonlinear optical response of the cavity material \cite{Ferretti2012}, or from polariton-polariton interactions in the case of exciton-polaritons \cite{Verger2006} or polaritons originating from driven few-level atoms \cite{Hartmann2008,Hartmann2007,Imamoglu1997,Werner1999}. In this second case, the resulting Hamiltonian is sometimes called the {\em Kerr-Hubbard (KH)} Hamiltonian, as it corresponds to the Bose-Hubbard Hamiltonian with the addition of driving fields and dissipation.

A vast theoretical literature has appeared in the last decade, describing several ways in which quantum phase transitions could be observed in coupled cavity arrays (as reviews, see Refs. \cite{Hartmann2008,Carusotto2012}). Early works \cite{Hartmann2006,Angelakis2007,Greentree2006,Hartmann2007,Chang2008} have focused on demonstrating how specific optical systems could be mapped onto the JCH or KH Hamiltonians, thus displaying typical physics of strongly correlated systems such as the quantum phase transition from a superfluid to a Mott insulator in cavity arrays, or the occurrence of the Tonks-Girardeau gas in an optical fiber. These works also suggested that the two models can coincide under specific conditions \cite{Greentree2006,Koch2009}. The case of disordered arrays was also considered \cite{Rossini2007}, where the quantum phase transition between a superfluid and a Bose-glass phase was characterized.

In these works however the driven-dissipative nature of the optical systems under investigation was only marginally taken into account, while an assumption of a lossless system at equilibrium was essentially employed to model the occurrence of quantum phase transitions. Within this assumption, the typical condition for the onset of strong quantum correlations is, analogously to the prototypical case of the Mott insulator, that $U\gg J$, where $U$ represents the on-site interaction energy per particle and $J$ the hopping energy between adjacent sites. It was only in subsequent works \cite{Verger2006,Gerace2009,Carusotto2009,Hartmann2010} that the important role of dissipation was clarified, showing that driven-dissipative systems may differ considerably from their lossless counterparts.

In a driven-dissipative system, each photon is confined in a cavity only during a time lapse corresponding to the cavity lifetime, after which it eventually radiates out of the cavity array. An external source -- typically one or more laser beams -- replenishes the system with new photons, and both time-dependent and steady-state regimes can be achieved depending on the experimental protocol. General wisdom suggests that, in a driven-dissipative system, quantum correlated states of many particles can be obtained only if the particle lifetime is long enough. More precisely, if one particle is added to the system, the time it takes to acquire quantum correlations with the other particles present in the array must be much shorter than the time the particle spends inside the system. The rates of these two processes are characterized by the effective on-site interaction energy per particle $U$, and by the loss rate $\gamma$ respectively. It is clear then, that the condition $U/\gamma\gg1$ is required in order for the many-particle state to acquire quantum correlations. The most illustrative example of this concept can be found already in the analysis of the single site KH problem, as was done for example in the context of polaritons in micropillars \cite{Verger2006}, where it is shown that the condition $U/\gamma\gg1$ is required in order to reach the blockade regime \cite{Imamoglu1997,Werner1999} and emit photons with sub-Poissonian statistics. The parameter $U/\gamma$ is an example of {\em cooperativity} \cite{Bowden2000} and it is indeed related to the atomic cooperativity parameter that enters the dissipative Jaynes-Cummings model of CQED \cite{Greentree2006} (although with a caveat \cite{Koch2009}) or more specific polaritonic models of few-level atoms \cite{Hartmann2008}. According to this intuitive argument, systems with small cooperativity should be restricted into non-entangled quantum states, generally having a classical analog.

Reaching high cooperativity in realistic systems is very challenging. Currently, the state-of-the-art value of this parameter is of the order of 1 in both atom-semiconductor-resonator \cite{Dayan2008} and all-semiconductor systems \cite{Reinhard2012}, while it reaches values of about 10 in atom-cavity systems \cite{Birnbaum2005} and of about 50 in superconducting circuits \cite{Lang2011}.

The high cooperativity argument however relies on a rather strong assumption, namely that at the initial time an extra particle added to the system is set into a separable state, namely is not already entangled to the other particles. While this would be the case for atomic gases, optical systems have as a unique feature that the source is a coherent field, with controlled amplitude and phase that are passed on to the many-photon system inside the cavity array. Quantum interference effects might then produce quantum correlations directly as a result of the driving process, in a time shorter than the lifetime. Following this general idea, we have recently shown that quantum effects such as sub-Poissonian statistics~\cite{Liew2010} can be expected in a system with very low cooperativity, under particular conditions. This is possible due to quantum interference between different excitation pathways and its sensitivity to small energy shifts caused by the interaction of pairs of particles~\cite{Bamba2011}. Theoretically, the quantum interference mechanism can also be exploited to create bipartite entangled states in a system of three coupled cavities~\cite{Liew2012}, where the entanglement is based on the amplitude and phases of quantum optical fields. This continuous variable (CV) entanglement \cite{Braunstein2005} can be detected using a homodyne interference setup \cite{Ou1992}, not being conditioned on single photon detection. Being issued of a quantum interference process, differently from the blockade effect, this mechanism is rather sensitive to the rate of pure dephasing characterising each quantum resonator. It is then very important to quantitatively account for pure dephasing in all model calculations.

A remaining question is whether multipartite entangled states can be produced in a larger array of linearly coupled optical resonators in the regime of low cooperativity. Besides the fundamental interest, multipartite entanglement represents an important resource for applications in quantum information science~\cite{Raussendorf2001,Menicucci2006}. While it is possible to produce multipartite states by coupling parametric down-conversion processes in nonlinear crystals~\cite{Pfister2004,Ferraro2004,Su2007,Yukawa2008,Midgley2010} or using optical frequency combs~\cite{Menicucci2008,Pysher2011}, arrays of optical cavities can be realized as compact solid-state systems, which are more appropriate for devices.

Here, after reviewing our analysis of a three-cavity array for the generation of two entangled photons \cite{Liew2012}, and extending it to the JCH Hamiltonian, we study middle-sized arrays of up to eight coupled modes. We demonstrate two schemes for the generation of multipartite CV-entanglement, which is assessed via the violation of van Loock-Furusawa inequalities~\cite{Loock2003} (a multipartite CV analogue of Bell inequalities). Each scheme is characterized by an array topology and by the selection of the subset of cavities that are not laser-driven in a mutually coherent fashion. Four-party CV-entanglement is achieved at an optimal amplitude of the driving field, corresponding to an average mode occupation of slightly below 1, and is tested against a realistic value of the pure dephasing rate. By modifying the relative phases of the driving fields, it is shown that some control of the entanglement is possible, holding promise for the realization of devices for the controlled generation of multipartite entangled photons. Previous theoretical calculations of small arrays have relied on the direct numerical solution of the Von Neumann equation for the density matrix in a truncated Fock space (particle number basis). While this is an accurate approach within quantum optics, it is numerically heavy and cannot be applied to systems of more than a few coupled cavities. In this work, we make use of the wavefunction Monte Carlo (WFMC) technique \cite{Molmer1993}, which gives access to the exact solution of the Von Neumann equation by trading off speed against low memory consumption.\\

\section{Theory}

We consider arrays of coupled nonlinear optical resonators under near-resonant excitation. The theoretical model will be suitable for the description of a variety of systems that can be divided into two types. First, polaritonic systems, in which the polaritons are well described as bosons with Bose-Hubbard-like interactions. This includes not only exciton-polaritons in semiconductor microcavities, but also the polaritons that arise whenever many two-level systems are resonantly coupled to an optical resonator \cite{Hartmann2008,Imamoglu1997,Werner1999}. Second, the case of a single few-level system (e.g., an atom or quantum dot) coupled to an optical resonator. In this latter case, the Kerr nonlinearity is a good description in the limiting case of the dispersive regime~\cite{Greentree2006,Koch2009}.

Under near-resonant excitation, higher energy modes in realistic systems can be neglected, such that in the model each resonator is described by a single mode with energy $E_j$. The cavities are also defined by a constant $U$ characterizing the nonlinear energy per particle, which we assume equal in each cavity for simplicity. A matrix $J_{jk}$ describes the tunnelling rates between pairs of resonators, which are related to the topology of the system (i.e., the spatial arrangement of the cavities). Under these assumptions the KH Hamiltonian~\cite{Hartmann2006} of the system is:

\begin{eqnarray}
\hat{\mathcal{H}}_S&=\sum_j\left[E_j\hat{a}^\dagger_j\hat{a}_j+U\hat{a}^\dagger_j\hat{a}^\dagger_j\hat{a}_j\hat{a}_j+F_j\left(\hat{a}_j+\hat{a}^\dagger_j\right)\right] \nonumber\\
&\hspace{30mm}+\sum_{jk}\left[J_{jk}\left(\hat{a}^\dagger_j\hat{a}_k+\hat{a}^\dagger_k\hat{a}_j\right)\right],\label{eq7}
\end{eqnarray}
where $\hat{a}_j$ are the Bose annihilation operators associated with the modes and $F_j$ are the complex amplitudes of driving fields originating from the optical pumps acting on the modes. Note that this Hamiltonian is written directly in the rotating frame of the driving field, so that $F_j$ is a constant in time and the energies $E_j$ are expressed relative to the optical pump energy $\hbar\omega_0$. For simplicity, we are assuming monochromatic excitation, namely that the optical fields acting on each cavity have the same energy. If this condition is lifted, then typically the system does not admit a steady state solution even for stationary driving fields, The study of this situation lies beyond the scope of the present work.

The description of a driven-dissipative system requires the introduction of a coupling to an external reservoir. Here we assume two effects originating from this coupling. First, to each cavity mode a dissipation process at a rate $\gamma$ is associated, corresponding to the finite lifetime of a photon inside the cavity. Second, we also introduce a pure dephasing rate associated to each mode, at a rate $\Gamma$. Pure dephasing is the result of an interaction of the photon with the random environment of the bath without leakage of the photon out of the cavity (for example the exciton-phonon scattering mechanism in the case of exciton-polaritons in semiconductor microcavities \cite{Savona1997}). In terms of the density matrix of the system, the pure dephasing describes a decay of the off-diagonal terms of the density matrix without a corresponding decay of the diagonal ones. For simplicity, all modes are assumed physically similar, being characterized by common dissipation and dephasing rates. Dissipation and pure dephasing can be accounted for in the Von Neumann equation for the density matrix by introducing corresponding Lindblad terms \cite{Walls1985}, as detailed in the next section. We note that the weakly nonlinear regime considered in this work is characterized by $\gamma\gg U$ and $J\gg U$.

We will consider two different approaches for studying the dynamics of this driven-dissipative system: the direct numerical solution of the Von Neumann equation for the density matrix, and the Wavefunction Monte Carlo (WFMC) approach. The former is accepted as the most accurate within the framework of quantum optics, however is not feasible for large systems due to the exceedingly large size of the corresponding density matrix. The latter has been shown to give equivalent results and is suitable for larger systems \cite{Molmer1993}. We will provide numerical evidence of the equivalence between the two approaches when reviewing our recent result in the case of a system of three coupled modes \cite{Liew2012} where bipartite entanglement is expected. Afterwards, we will consider systems of seven and eight coupled cavities and demonstrate that they can give rise to multipartite entanglement of four photons under appropriate conditions. For these larger systems, we will restrict to the use of the WFMC approach, as a direct solution of the density matrix equations would be numerically unfeasible.

\subsection{Von Neumann Equation}\label{sec:MasterEquation}

The quantum optical behaviour of our system can be fully described using the Von Neumann equation for the density matrix, $\boldsymbol{\rho}$:

\begin{eqnarray}
i\hbar\frac{d \boldsymbol{\rho}}{dt}&=\left[\hat{\mathcal{H}_S},\boldsymbol{\rho}\right]+i\frac{\gamma}{2}\sum_j\left(2\hat{a}_j\boldsymbol{\rho}\hat{a}^\dagger_j-\hat{a}^\dagger_j\hat{a}_j\boldsymbol{\rho}-\boldsymbol{\rho}\hat{a}^\dagger_j\hat{a}_j\right)\nonumber\\
&\hspace{30mm}+i\frac{\Gamma}{2}\sum_j\left(2\hat{n}_j\boldsymbol{\rho}\hat{n}_j-\hat{n}^2_j\boldsymbol{\rho}-\boldsymbol{\rho}\hat{n}^2_j\right),\label{eq:master}
\end{eqnarray}
The second and third terms on the right-hand side of this equation are Lindblad terms required to describe the Markovian coupling between the system and a random environment. They describe respectively the dissipation and pure dephasing processes \cite{Walls1985}. Here, $\hat{n}_j=\hat{a}_j^\dagger\hat{a}_j$ is the occupation operator of the $j$-th mode.
Equation~\ref{eq:master} can be solved by expanding the density matrix over a particle
number basis in a similar way to that done in Ref.~\cite{Verger2006}:

\begin{equation}
\boldsymbol{\rho}=\left|n_1,n_2,...,n_N\right>\left<n_1^\prime,n_2^\prime,...,n_N^\prime\right|\rho_{n_1,n_2,...,n_N,n_1^\prime,n_2^\prime,...,n_N^\prime}\label{eq:rhoexpand}
\end{equation}
It is necessary to truncate at a given particle number, $n_i<n_{max}$, which is accurate in the regime of small occupations. Substitution of Eq.~\ref{eq:rhoexpand} into Eq.~\ref{eq:master} gives a finite set of evolution equations for the elements of the density matrix, $\rho_{n_1,n_2,...,n_N,n_1^\prime,n_2^\prime,...,n_N^\prime}$ (see the supplemental material of Ref.~\cite{Liew2010} for more detail). One can then time evolve the system numerically from the vacuum to obtain the steady state density matrix, from which all observable quantities can be calculated. It is often easier however to solve directly for the steady state -- when assuming constant $F_j$'s -- by solving the linear homogeneous set of equations obtained for $d\boldsymbol{\rho}/dt=0$, with the additional condition $\mbox{tr}(\boldsymbol{\rho})=1$. As an additional simplification of the numerical calculations, it is straightforward to further truncate the basis of the wavefunction such that only states $\left|n_1,n_2,...,n_N\right>$ with $\sum_mn_m\leq n_{max}$ are included. For the same value of $n_{max}$, this prescription amounts to selecting a considerably smaller Hilbert space. The justification of this choice lies in the fact that this latter truncation scheme retains all states within a maximum total energy, responding to an intuitive perturbation criterion, while the former one includes many more states all characterized by higher energy, which are presumably less relevant in a perturbation expansion.

\subsection{Wavefunction Monte Carlo Technique}\label{sec:WFMC}

The WFMC technique was developed and reported in Ref.~\cite{Molmer1993}. Here we give a brief summary of the basic elements applied to the class of systems we are investigating. Time evolution of the density matrix using the Von Neumann equation is replaced with time evolution of a wavefunction, $\left|\psi(t)\right>$, using the non-Hermitian Hamiltonian
\begin{equation}
\hat{\mathcal{H}}=\hat{\mathcal{H}}_S-\frac{i\hbar}{2}\gamma\sum_n\hat{n}_n-\frac{i\hbar}{2}\Gamma\sum_n \hat{n}_n^2,
\end{equation}
where $\hat{n}_n=\hat{a}^\dagger_n\hat{a}_n$. Since this Hamiltonian is non-Hermitian it is important to renormalize the wavefunction at every timestep in a numerical calculation such that the normalization condition $\left<\psi(t)|\psi(t)\right>=1$ is preserved during the time evolution.

In addition to evolution with the non-Hermitian Hamiltonian, there is a possibility of a stochastic quantum jump in the wavefunction between the times $t$ and $t+\delta t$. Two different types of jumps are possible, associated with either the dissipation or pure dephasing:
\begin{eqnarray}
\left|\psi(t+\delta t)\right>&=\hat{a}_m \left|\psi(t)\right>,\label{eq:jump1}\\
\left|\psi(t+\delta t)\right>&=\hat{n}_m \left|\psi(t)\right>,\label{eq:jump2}
\end{eqnarray}
which occur, respectively with probabilities
\begin{eqnarray}
\delta p_m(t)&=\gamma\delta t\left<\psi(t)|\hat{n}_m|\psi(t)\right>\\
\delta p_m^\prime(t)&=\Gamma\delta t\left<\psi(t)|\hat{n}_m^2|\psi(t)\right>.
\end{eqnarray}
Using these probabilities, the WFMC procedure is equivalent to the simulation of a stochastic process. By repeating the calculation over a large enough ensemble of time sequences of stochastic quantum jumps, a distribution of different wavefunctions can be obtained at each time. Note that one should choose a short enough timestep such that $\sum_m\left(\delta p_m+\delta p_m^\prime\right)\ll1$. It is also important to renormalize the wavefunction after a jump, since Eqs.~\ref{eq:jump1}-\ref{eq:jump2} do not preserve the normalization.

In Ref.~\cite{Molmer1993} it was shown that the WFMC procedure is equivalent to the Von Neumann equation approach when observable quantities are calculated by averaging over multiple wavefunction realizations. The expectation value of the operator $\hat{\mathcal{O}}$ is given by averaging over $N_R$ different realizations of the wavefunction $\left|\psi^{(i)}(t)\right>$, each obtained from a different time sequence of quantum jumps:
\begin{equation}
\left<\hat{\mathcal{O}}\right>=\frac{1}{N_R}\sum_{i=1}^{N_R}\left<\psi^{(i)}(t)|\hat{\mathcal{O}}|\psi^{(i)}(t)\right>\label{eq:expectation}
\end{equation}
In this paper, the systems that we study admit steady state solutions. After a steady-state is reached one can also perform time averaging, which reduces the statistical error of the technique.

The advantage of the WFMC technique is that the wavefunction is expanded over a basis with less than $(n_{max}+1)^M$ states -- where $M$ is the number of cavities -- in contrast to the direct solution of the equation for the density matrix where $(n_{max}+1)^{2M}$ equations need to be solved. This enables the solution of problems that would otherwise require amounts of memory too large to be handled with standard modern computers. As for the direct solution of the master equations, one can reduce the size of the basis by including only states with $\sum_mn_m\leq n_{max}$. In all numerical solutions of the problem presented below, we have carefully checked the convergence of the result as a function of $n_{max}$. Typically, convergence was reached for values of $n_{max}$ ranging from 8 to 12 depending on the array and parameter range under study.

\section{Two Party Entanglement in a Three Coupled Mode System}

\begin{figure}[h]
\centering
\includegraphics[width=10cm]{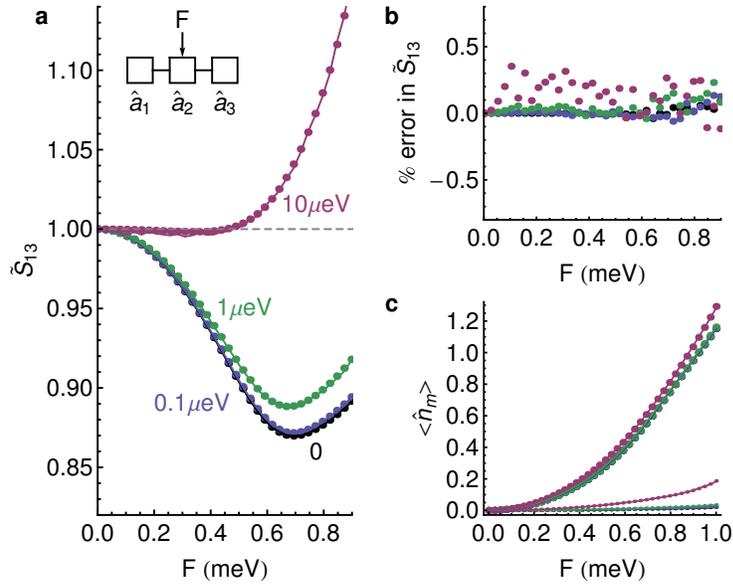}
\caption{(color online) a) Variation of the parameter $\tilde{S}_{13}$ with driving field amplitude for a chain of three coupled cavities. Spots correspond to results from the Von Neumann equation approach, while curves correspond to results from the WFMC approach. Different traces (spots and curves) correspond to different values of pure dephasing, $\Gamma$ (values marked on plot). The inset shows the topology of the cavity array, where the central cavity is driven. b) \% error of WFMC approach. c) Variation of the occupation numbers of the cavities with driving field amplitude. Again, spots correspond to results from the Von Neumann equation approach, while curves correspond to results from the WFMC approach. The large spots correspond to the average occupation of the modes $\langle \hat{n}_1\rangle=\langle \hat{n}_3\rangle$, while the small spots correspond to $\langle \hat{n}_2\rangle$. Parameters: $U=0.012$meV, $\gamma=0.044$meV, $E_1+E_3=-0.06$meV ($E_1=E_3$ in a and b), $E_2=0.08$meV.}
\label{fig:ThreeModes}
\end{figure}

In this section we review our recent result \cite{Liew2012} on a system of three cavities coupled in a chain, where the central cavity is driven by a near-resonant optical excitation (see inset in Fig.~\ref{fig:ThreeModes}a). This system is sufficiently simple to be treated using both the full Von Neumann equation approach (section~\ref{sec:MasterEquation}) and the WFMC approach (section~\ref{sec:WFMC}), which allows to test the effectiveness of the WFMC method. In the system under consideration, it has been previously shown that two-party CV entanglement~\cite{Braunstein2005} can develop between the two non-driven modes due to the presence of quantum interference in the system~\cite{Liew2012}.

In analogy to Bell's result for discrete variable entanglement, CV entanglement between modes $n$ and $m$ is characterized by the violation of an inequality~\cite{Duan2000,Simon2000}:
\begin{equation}
1\leq S_{nm}=V\left(\hat{p}_n-\hat{p}_m\right)+V\left(\hat{q}_n+\hat{q}_m\right),\label{eq:S}
\end{equation}
where the amplitude and phase operators are defined as $\hat{p}_n=\left(\hat{a}_n+\hat{a}^\dagger_n\right)/2$ and $\hat{q}_n=\left(\hat{a}_n-\hat{a}^\dagger_n\right)/(2i)$, respectively. The variance of an operator is defined by $V(\hat{\mathcal{O}})=\langle\hat{\mathcal{O}}^2\rangle-\langle\hat{\mathcal{O}}\rangle^2$. To be sure that inequality~\ref{eq:S} evidences entanglement if it is present, it may be necessary to minimize the value of $S_{13}$ over the phase references of the different modes. Local operations allow to make the transformation $\hat{a}_n\mapsto\hat{a}_ne^{-i\phi_n}$ and the minimum value of $S_{nm}$, obtained when $\phi_m+\phi_n$ is an integer multiple of $2\pi$, is given by:
\begin{eqnarray}
\tilde{S}_{nm}&=1+\langle\hat{a}^\dagger_n\hat{a}_n\rangle+\langle\hat{a}^\dagger_m\hat{a}_m\rangle-\langle\hat{a}^\dagger_n\rangle\langle\hat{a}_n\rangle-\langle\hat{a}^\dagger_m\rangle\langle\hat{a}_m\rangle\nonumber\\
&-2\sqrt{\langle\hat{a}^\dagger_n\hat{a}^\dagger_m\rangle-\langle\hat{a}^\dagger_n\rangle\langle\hat{a}^\dagger_m\rangle}\sqrt{\langle\hat{a}_n\hat{a}_m\rangle-\langle\hat{a}_n\rangle\langle\hat{a}_m\rangle}\label{eq:Sopt}
\end{eqnarray}
The expectation values entering this equation can be calculated from the density matrix, $\boldsymbol{\rho}$, or from Eq.~(\ref{eq:expectation}), depending on the theoretical approach used to model the system.

The correlation (\ref{eq:S}) is evaluated at equal times. To measure this kind of correlation then, the balanced homodyne technique should be best suited \cite{Braunstein2005}. It should be noted however, that in this system two-times correlation functions are expected to vary as a function of the time delay over a scale defined by the inverse of the tunneling rate $J$, which is assumed to be the largest energy scale in the system, similarly to the two-cavity system studied in Ref. \cite{Liew2010}. A measurement should then either rely on photodetectors with short enough time resolution, or employ a pulsed homodyne scheme \cite{Lvovsky2009}, where the time resolution is set by the pulse duration of the local oscillator independently of the time-resolution of the detectors.

The variation of $\tilde{S}_{13}$ with driving field amplitude is shown in Fig.~\ref{fig:ThreeModes}, using the same set of parameters as in Ref.~\cite{Liew2012}. These parameters were originally chosen to correspond to exciton-polariton boxes~\cite{Verger2006} of size $3\mu m$ separated by $1\mu m$, for which $J=0.5$meV~\cite{Liew2010} can be obtained from solution of the Schr\"odinger equation in a two well potential (this value is also in agreement with experimental measurements~\cite{Vasconcellos2011,Galbriati2012}). It is well-known how to calculate the nonlinear interaction strength~\cite{Verger2006} and we take the value $U=0.012$meV in agreement with recent experimental data~\cite{Kasprzak2007,Amo2009,Ferrier2011}. The exciton-polariton decay rate was taken as $\gamma=0.044$meV, which was measured in Ref.~\cite{Wertz2010} and we chose $E_1+E_3=-0.06$meV ($E_1=E_3$ in a and b), $E_2=0.08$meV. The slight detuning between the modes $E_{1,3}$ and $E_2$ was previously found to give the smallest value of $\tilde{S}_{13}$ for fixed $J$ and $\gamma$ in Ref.~\cite{Liew2012}. It should be noted that the master equations issued of the Hamiltonian (\ref{eq7}), with dissipation and dephasing rates $\gamma$ and $\Gamma$ respectively, can be reduced in a dimensionless form by measuring all energy parameters with respect to $\gamma$. Hence, all results shown here are invariant under rescaling of all energies and therefore hold for systems characterized by different ranges of parameters such as, for example, superconducting devices \cite{Lang2011}.

In agreement with the calculations based on the Von Neumann equations (spots in Fig.~\ref{fig:ThreeModes}), the WFMC approach (curves) demonstrates entanglement between the first and third modes in the system via violation of the inequality $1\leq \tilde{S}_{13}$. The differences between the two approaches are shown in Fig.~\ref{fig:ThreeModes}b, which illustrates that the WFMC approach is an accurate alternative technique for our problem. It is notable that, when applying MCWF, larger dephasing rates require averaging over larger Monte Carlo samples in order to achieve the desired accuracy. However, since realistic estimates of the dephasing rate in semiconductor micropillars are in the range of tenths of $\mu$eV~\cite{Savona1997}, the technique turns out to be very powerful for the purposes of this paper. A minimum in the value of $\tilde{S}_{13}$ is obtained for a driving rate of $F\simeq0.7~\mbox{meV}$, while for larger rates this quantity rapidly increases to values above 1, thus recovering a non-entangled state as expected in the limit of high average occupation of the modes. The variation of the occupation numbers with driving field amplitude is shown in Fig.~\ref{fig:ThreeModes}c, where again there is full agreement between the two approaches. It is important to notice that entanglement occurs for average occupation below unity, where nonclassical effects are indeed expected. As illustrated in Ref. \cite{Liew2012}, this entanglement can be characterized as a deviation from a Glauber state (thus separable) of the sum of modes 1 and 3. The deviation consists in an increase of the amplitude of Fock states $|N,N^\prime,0\rangle$ and $|0,N^\prime,N\rangle$ which, in the subspace of modes 1 and 3, corresponds to an increased amplitude of the maximally entangled states $2^{-1/2}(|N,0\rangle+|0,N\rangle)$.
\begin{figure}[h]
\centering
\includegraphics[width=10cm]{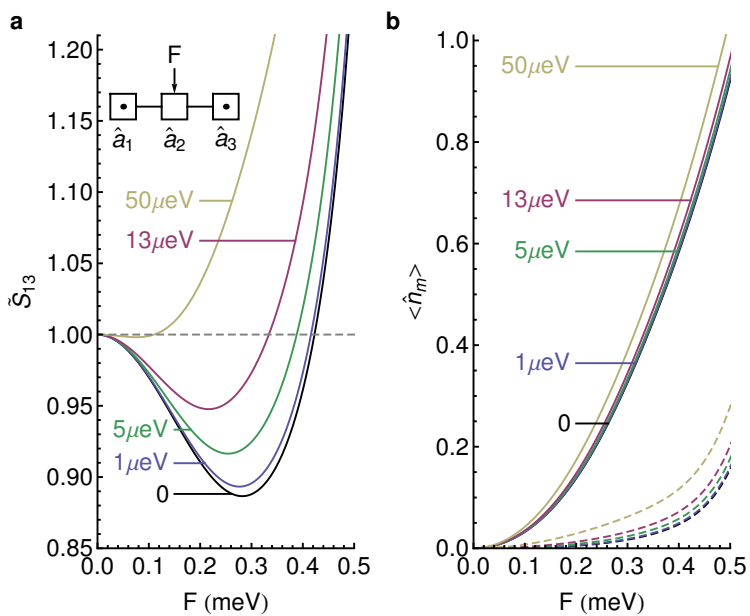}
\caption{(color online) a) Variation of the parameter $\tilde{S}_{13}$ with driving field amplitude for a chain of three coupled cavities, in the case of the JCH model. Different traces correspond to different values of pure dephasing acting on the two-level systems, $\Gamma_s$ (values marked on plot). b) Variation of the occupation numbers of cavities 1 and 3 (full lines) and of cavity 2 (dashed lines) with driving field amplitude. Different colors denote different values of $\Gamma_s$ as in panel a). Parameters: $g=141~\mu\mbox{eV}$, $\gamma=53~\mu\mbox{eV}$, $E_1=E_3=0.06~\mbox{meV}$, $E_2=-0.12~\mbox{meV}$, $E_{1s}=E_{3s}=0.22~\mbox{meV}$, $J=0.29~\mbox{meV}$.}
\label{fig:JCH}
\end{figure}

To conclude this section on bipartite entanglement, we extend our previous result \cite{Liew2012} by modelling the same array geometry in which the Kerr-Hubbard Hamiltonian is replaced by a Jaynes-Cummings-Hubbard one. The question whether the present mechanism holds in the JCH case is a nontrivial one, as the JCH Hamiltonian is not in general equivalent to a KH one \cite{Greentree2006,Koch2009}. In the limit of very large detuning between the cavity and the two-level system, the JCH case can indeed be associated to an effective KH Hamiltonian. The case of major relevance however is the one close to resonance, as in all atom-cavity, semiconductor-QD, or Josephson-qubit systems. The weak cooperativity limit was already studied in the case of two coupled cavities, one of which coupled to a two-level system \cite{Bamba2011}, showing that an effect quantitatively analogous to the one arising in the KH case can be obtained for moderate cooperativity ($g/\gamma\simeq 1$, where $g$ is the Jaynes-Cummings coupling constant introduced below). Here we study the three-cavity configuration, assuming that each of the two side ones is coupled to a two-level system. The Hamiltonian of this model system is
\begin{eqnarray}
\hat{\mathcal{H}}_S&=&\sum_{j=1}^3E_j\hat{a}^\dagger_j\hat{a}_j+F\left(\hat{a}_2+\hat{a}^\dagger_2\right)+E_{1s}\hat{\sigma}_{1+}\hat{\sigma}_{1-}+E_{3s}\hat{\sigma}_{3+}\hat{\sigma}_{3-} \nonumber\\
&+&g\left(\hat{\sigma}_{1+}\hat{a}_1+\hat{\sigma}_{1-}\hat{a}_1^\dagger\right)
+g\left(\hat{\sigma}_{3+}\hat{a}_3+\hat{\sigma}_{3-}\hat{a}_3^\dagger\right) \nonumber\\
&+&J\left(\hat{a}^\dagger_1\hat{a}_2+\hat{a}^\dagger_2\hat{a}_1\right)+J\left(\hat{a}^\dagger_3\hat{a}_2+\hat{a}^\dagger_2\hat{a}_3\right),\label{JCH}
\end{eqnarray}
where $\hat{\sigma}_{j\pm}$ are the raising and lowering operators of the $j$-th two-level system respectively. Lindblad terms for the two-level systems are written analogously to those for the cavity modes. Here, in addition to taking a single value for $J$ and for $g$, we also assume degenerate side cavities, namely $E_1=E_3$ and $E_{1s}=E_{3s}$. For the numerical calculations, we take a dissipation rate of the two-level systems $\gamma_s=\gamma/100$. This value should model nonradiative decay channels as well as decay into photonic modes other than the cavity one, and matches typical values extrapolated for semiconductor systems \cite{Khitrova2006}. We neglect pure dephasing processes for the three cavity modes and retain only a pure dephasing rate $\Gamma_s$ for the two-level systems, as this process is expected to be the dominant source of dephasing. Fig.~\ref{fig:JCH}a shows the variation of the parameter $\tilde{S}_{13}$ with driving field amplitude, computed as in the KH case, while in  Fig.~\ref{fig:JCH}b the corresponding average occupations are plotted. Here we assumed parameters as in the system of a quantum dot in a photonic crystal cavity studied in Ref. \cite{Reinhard2012}, namely $g=141~\mu\mbox{eV}$ and $\gamma=53~\mu\mbox{eV}$. We have run a minimum search procedure on the energies and found an optimal violation of the generalized Bell's inequality for values $E_1=E_3=0.06~\mbox{meV}$, $E_2=-0.12~\mbox{meV}$, $E_{1s}=E_{3s}=0.22~\mbox{meV}$, and $J=0.29~\mbox{meV}$, as shown in  Fig.~\ref{fig:JCH}a. At the optimal value of $F$ the violation is of the same order as in the KH case, showing that the mechanism studied in this work does not necessarily rely on a specific kind of nonlinearity. Curves computed for the same values of the parameters but in presence of a finite $\Gamma_s$ are also shown. The pure dephasing of an InGaAs quantum dot has been extensively characterized by coherent four-wave mixing measurements, and its rate typically amounts to $\Gamma_s=1~\mu\mbox{eV}$ \cite{Borri2001}, or even smaller \cite{Langbein2004}. For this value we see that the value of $\tilde{S}_{13}$ is only slightly affected and the non-separability of the steady state is robust to pure dephasing rates up to ten times larger. Coupling between photonic crystal cavities is currently feasible and can be easily tuned \cite{Sato2012}. This result thus suggests that current semiconductor technology would be advanced enough to produce a proof-of-principle device for bipartite continuous-variable entangled photon generation.

\section{Multiparty Entanglement}

\begin{figure}[h]
\centering
\includegraphics[width=10cm]{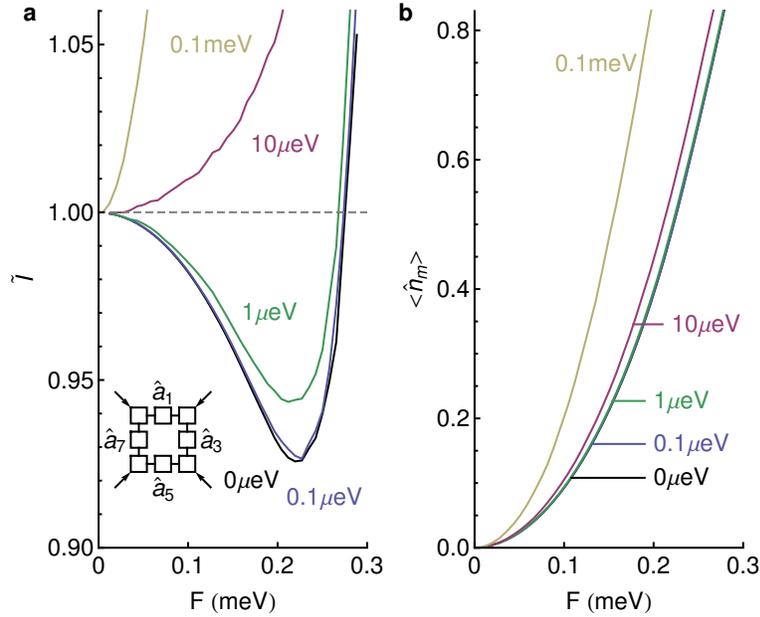}
\caption{(color online) a) Variation of the quadripartite entanglement parameter $\tilde{I}$ with driving field amplitude in the case of a ring of eight coupled modes with near-resonant optical pumping of alternate modes (illustrated by the inset). The different curves correspond to different values of pure dephasing (marked on the plot). b) Variation of the average occupation of the modes (note that all modes have the same average occupation due to symmetry). Parameters: $\Gamma=0.044$meV, $U=0.012$meV, $J=0.16$meV. The mode detunings with respect to the pump energy were chosen to minimize $\tilde{I}$: $E_{\mathrm odd}=-0.03$meV, $E_{\mathrm even}=0.11$meV.}
\label{fig:Quadripartite}
\end{figure}

The scheme discussed in the previous section can be generalized to many resonators. The objective is the generation of cluster-entangled states, that are a universal resource for quantum information processing \cite{Raussendorf2001}. In very simple terms, a cluster state is a state of many qubits arranged on an array, such that each qubit is entangled with the neighbouring ones. It is then natural to expect the emergence of a cluster state from the spatial repetition of the three cavity system discussed above. Hence, as a first example of an array topology able of generating such a multipartite entangled state, we consider eight cavities arranged on a ring, with next neighbour coupling, as illustrated in the inset to Fig.~\ref{fig:Quadripartite}a. Alternate cavities are optically driven, so as to repeat the three-cavity pattern of the previous section. Quadripartite entanglement should build up between the four non-driven modes. For simplicity we assume that the driving fields have equal amplitudes, such that: $F_{\mathrm{even}}=F$ and $F_\mathrm{odd}=0$. As before, we compute the steady state solution of the equations for the density matrix, issued from the Hamiltonian (\ref{eq7}), in presence of dissipation and pure dephasing.

Quadripartite continuous variable entanglement can be evidenced by the simultaneous violation of three inequalities~\cite{Midgley2010}:
\begin{eqnarray}
I_a=V(\hat{p}_1-\hat{p}_2)+V(\hat{q}_1+\hat{q}_2+g_A \hat{q}_3+g_B \hat{q}_0)&\geq1,\label{eq:ineq1}\\
I_b=V(\hat{p}_2-\hat{p}_3)+V(g_C \hat{q}_1+\hat{q}_2+\hat{q}_3+g_B \hat{q}_0)&\geq1,\label{eq:ineq2}\\
I_c=V(\hat{p}_3-\hat{p}_0)+V(g_C \hat{q}_1+g_D \hat{q}_2+\hat{q}_3+\hat{q}_0)&\geq1,\label{eq:ineq3}
\end{eqnarray}
where $\{g_A, g_B, g_C, g_D\}$ are arbitrary real parameters that should be chosen to minimize the largest value of the set of $\{I_a, I_b, I_c\}$, that is, $\tilde{I}=\min_{g_i}\left(\max\{I_a,I_b,I_c\}\right)$. As in the bipartite case, an additional optimization is made by varying the phase reference of each of the modes. We note that the quantities $\{I_a, I_b, I_c\}$ can be re-written:
\begin{eqnarray}
I_a&=V_1^\prime+V_2^\prime-2V_{12}^\prime+V_1+V_2+2V_{12}+2g_A\left(V_{13}+V_{23}\right)+2g_B\left(V_{01}+V_{02}\right)\nonumber\\
&\hspace{5mm}+g_A^2V_3+2g_Ag_B V_{03}+g_B V_0,\\
I_b&=V_2^\prime+V_3^\prime-2V_{23}^\prime+V_2+V_3+2V_{23}+2g_B\left(V_{02}+V_{03}\right)+2g_C\left(V_{12}+V_{13}\right)\nonumber\\
&\hspace{5mm}+g_B^2V_0+2g_Bg_C V_{01}+g_C V_1,\\
I_c&=V_3^\prime+V_0^\prime-2V_{03}^\prime+V_3+V_0+2V_{03}+2g_C\left(V_{13}+V_{01}\right)+2g_D\left(V_{23}+V_{02}\right)\nonumber\\
&\hspace{5mm}+g_C^2V_1+2g_Cg_D V_{12}+g_D V_2,
\end{eqnarray}
where:
\begin{eqnarray}
V_{nm}&=\left<\hat{q}_n\hat{q}_m\right>-\left<\hat{q}_n\right>\left<\hat{q}_m\right>\\
V_{n}&=\left<\hat{q}_n^2\right>-\left<\hat{q}_n\right>^2\\
V^\prime_{nm}&=\left<\hat{p}_n\hat{p}_m\right>-\left<\hat{p}_n\right>\left<\hat{p}_m\right>\\
V^\prime_{n}&=\left<\hat{p}_n^2\right>-\left<\hat{p}_n\right>^2
\end{eqnarray}
Due to the symmetry of the system made by a ring of cavities (inset to Fig.~\ref{fig:Quadripartite}a), the minimum value of $\tilde{I}$ is given when $I_a=I_b=I_c$ and $g_A=g_B=g_C=g_D$. We then obtain:
\begin{equation}
g_{A,B,C,D}=\frac{V_{01}+V_{02}}{V_0+V_{01}}
\end{equation}
Results from the WFMC approach are shown in Fig.~\ref{fig:Quadripartite}a, where the variation of $\tilde{I}$ with driving field amplitude is shown for a range of values of pure dephasing. For small dephasing, the value of $\tilde{I}$ drops below $1$ indicating the presence of quadripartite entanglement between the modes $\hat{a}_1$, $\hat{a}_3$, $\hat{a}_5$, and $\hat{a}_7$. This entanglement survives realistic values of pure dephasing in the range of tenths of $\mu$eV~\cite{Savona1997}. The entanglement is lost for large driving field amplitude, as before, because the system approaches a classical behaviour for large average occupation of the modes. For the optimum value of $\tilde{I}$, the average occupations of the modes, shown in Fig.~\ref{fig:Quadripartite}b, are approximately $0.45$ each, again well in the quantum regime.

\begin{figure}[h]
\centering
\includegraphics[width=10cm]{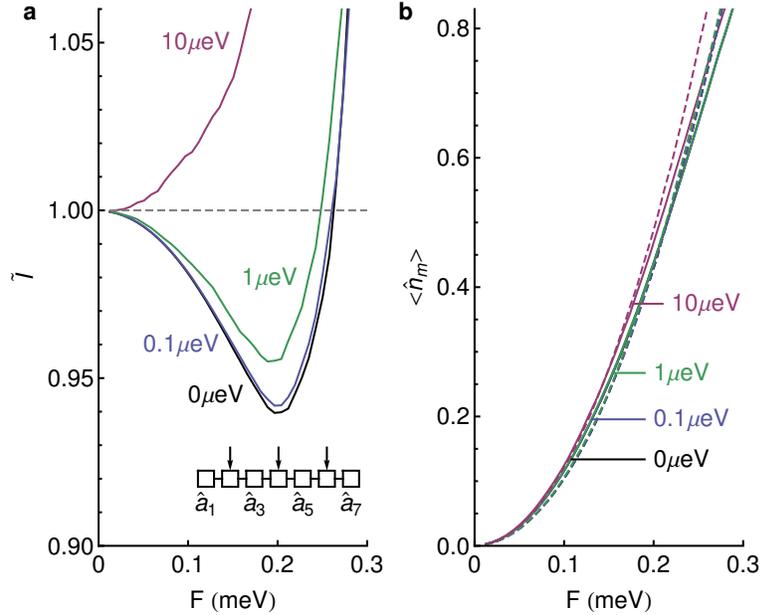}
\caption{(color online) a) Same as in Fig.~\ref{fig:Quadripartite} for a linear chain of modes (illustrated by the inset in a). The mode detunings with respect to the pump energy were chosen to minimize $\tilde{I}$: $E_{\mathrm odd}=-0.02$meV, $E_{\mathrm even}=-0.19$meV. The other parameters were the same as in Fig.~\ref{fig:Quadripartite}. In (b) the solid curves show the values of $\langle\hat{n}_1\rangle=\langle\hat{n}_7\rangle$, while the dashed curves show the values of $\langle\hat{n}_3\rangle=\langle\hat{n}_5\rangle$ (these mode occupations are equal due to symmetry).}
\label{fig:QuadripartiteLinear}
\end{figure}

Figure~\ref{fig:QuadripartiteLinear} shows an alternative scheme of operation based on a chain of seven modes (illustrated in the inset to Fig.~\ref{fig:QuadripartiteLinear}a). In this case the minimum value of $\tilde{I}$ does not occur for $g_A=g_B=g_C=g_D$. Instead, we adopt the scheme of Ref.~\cite{Midgley2010} where $g_A$ and $g_B$ are chosen to minimize $I_a$ while $g_C$ and $g_D$ are chosen to minimize $I_c$:
\begin{eqnarray}
g_A&=\frac{V_0\left(V_{13}+V_{23}\right)-V_{03}\left(V_{01}+V_{02}\right)}{V_{03}^2-V_0V_3}\\
g_B&=\frac{V_3\left(V_{01}+V_{02}\right)-V_{03}\left(V_{13}+V_{23}\right)}{V_{03}^2-V_0V_3}\\
g_C&=\frac{V_2\left(V_{01}+V_{13}\right)-V_{12}\left(V_{02}+V_{23}\right)}{V_{12}^2-V_1V_2}\\
g_D&=\frac{V_1\left(V_{02}+V_{23}\right)-V_{12}\left(V_{01}+V_{13}\right)}{V_{12}^2-V_1V_2}
\end{eqnarray}
One may worry that this procedure would not give a minimum value of $I_b$, however, for the states considered we found that $I_b<I_a$ and $I_b<I_c$. A numerical minimization procedure of $\tilde{I}$ was also considered and gave results matching those from the procedure of Ref.~\cite{Midgley2010}. Finally, as for the $S_{nm}$ parameter in the bipartite case, pulsed homodyne detection should be the best suited technique to measure these equal-time correlations.

Similar to the case in Fig.~\ref{fig:Quadripartite}, driving alternate modes allows quadripartite entanglement of the non-driven modes. This demonstrates the potential of arrays of weakly nonlinear cavities to generate entanglement in a variety of topologies that can be physically engineered. Again, the optimum value of $\tilde{I}$ corresponds to an average mode occupation of approximately $0.45$ (see Fig.~\ref{fig:QuadripartiteLinear}b). As in the three-mode case, entanglement here can be pictured as an increased amplitude of a four-mode cluster state with respect to the separable state that would arise for vanishing nonlinearity.

\begin{figure}[h]
\centering
\includegraphics[width=10cm]{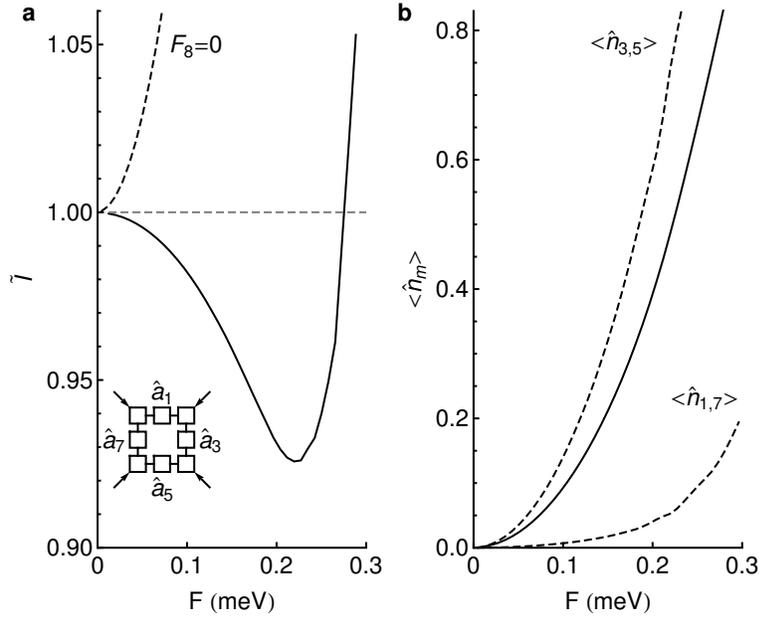}
\caption{(color online) a) The effect of removing one of the driving fields in the eight mode ring scheme is shown by the dashed curve (the solid curve is the same as in Fig.~\ref{fig:Quadripartite})a. b) The variation of the average mode occupations with all eight driving fields (solid curve) and one driving field removed (dashed curves). Due to symmetry, $\langle\hat{n}_3\rangle=\langle\hat{n}_5\rangle$ and $\langle\hat{n}_1\rangle=\langle\hat{n}_7\rangle$ when the driving field $F_8$ is removed. The other parameters were the same as in Fig.~\ref{fig:Quadripartite}a, with $\Gamma=0$.}
\label{fig:QuadripartiteOnePumpMissing}
\end{figure}

It is possible to switch off the quadripartite entanglement by changing the configuration of the driving fields. Removal of one of the driving fields in the eight mode system ($F_8=0$) causes the loss of quadripartite entanglement as shown by the dashed curve in Fig.~\ref{fig:QuadripartiteOnePumpMissing}a. Note that this system, with one driving field removed, is not identical to the chain of seven modes considered in Fig.~\ref{fig:QuadripartiteLinear}. With one driving field removed, and no re-optimization of parameters, the occupations of the different modes in the system may be very different, as shown in Fig.~\ref{fig:QuadripartiteOnePumpMissing}b. Indeed, when the mode occupations are not similar, one does not expect a strong entanglement since there is a dominant probability of detecting particles in a subset of modes.

\begin{figure}[h]
\centering
\includegraphics[width=10cm]{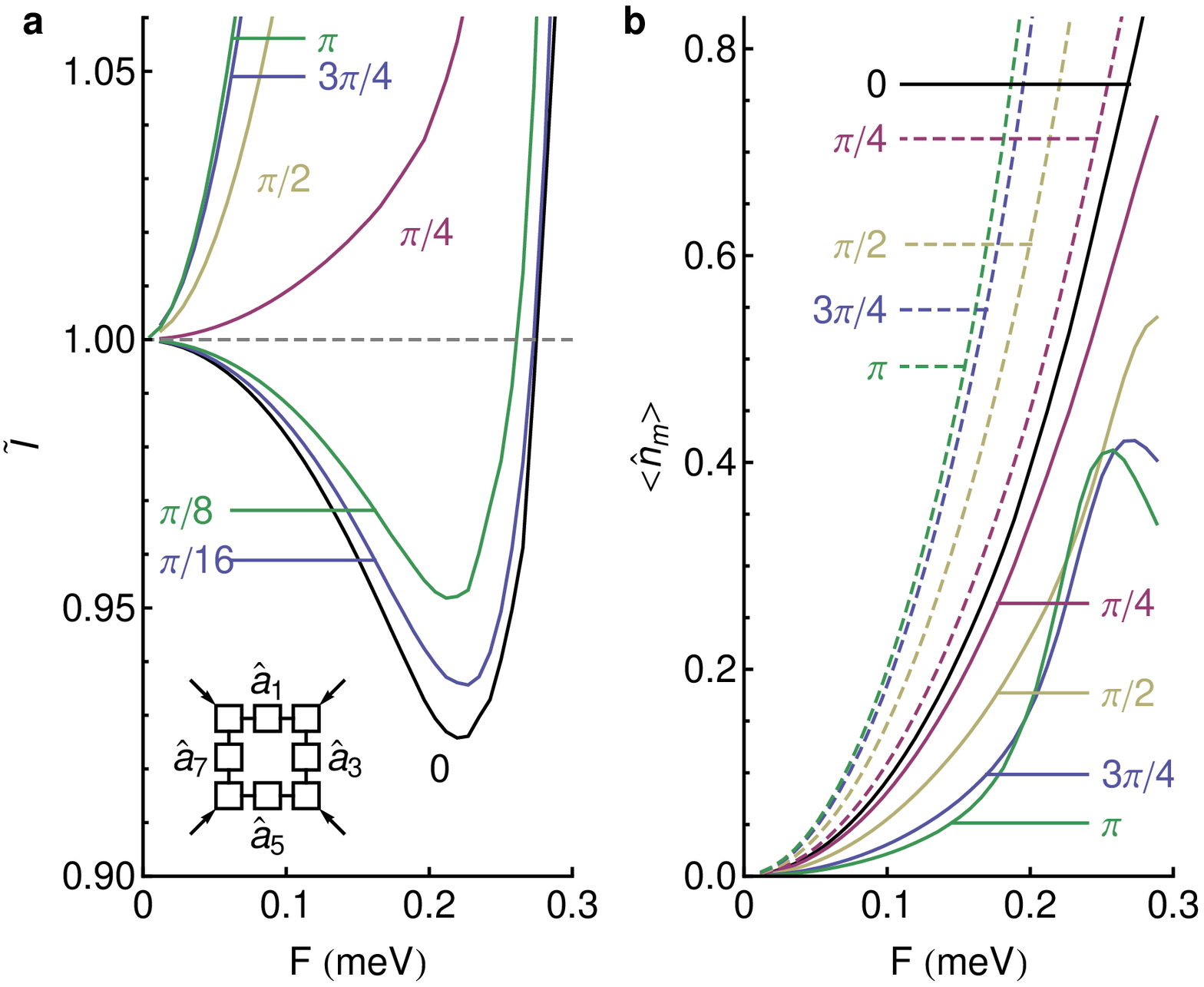}
\caption{(color online) The effect of varying the phase of one of the driving fields (numbers marked on the plot indicate the relative phases used). In (b) the solid curves show the values of $\langle\hat{n}_1\rangle=\langle\hat{n}_7\rangle$, while the dashed curves show the values of $\langle\hat{n}_3\rangle=\langle\hat{n}_5\rangle$ (these mode occupations are equal due to symmetry). Parameters the same as in Fig.~\ref{fig:Quadripartite}a, with $\Gamma=0$.}
\label{fig:QuadripartiteVariations}
\end{figure}

In addition, the entanglement is sensitive to the relative phase of the driving fields. Keeping all fields switched on but varying the phase of one of the fields ($F_8$) is also able to switch off the entanglement, as shown in Fig.~\ref{fig:QuadripartiteVariations}a where the entanglement parameter is plotted for a range of relative phases. Again, there is an associated imbalance of average mode occupations when varying the phase of one of the fields as shown in Fig.~\ref{fig:QuadripartiteVariations}b.

\begin{figure}[h]
\centering
\includegraphics[width=10cm]{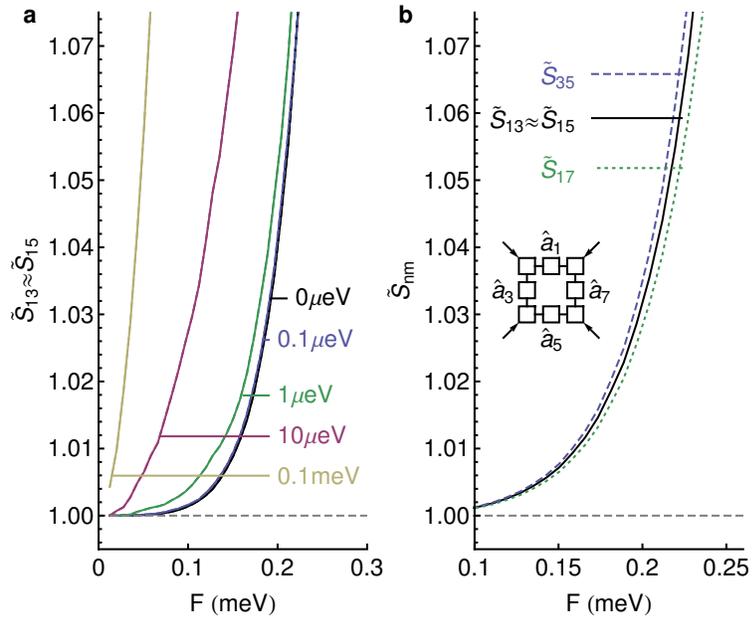}
\caption{(color online) a) Variation of the bipartite parameter $\tilde{S}_{13}$ with driving field amplitude for the eight mode ring for different values of pure dephasing. Note that $\tilde{S}_{13}\approx \tilde{S}_{15}$ and other combinations of modes are equivalent to either $\tilde{S}_{13}$ or $\tilde{S}_{15}$ due to symmetry. b) Variation of the bipartite parameters in the eight mode system with $F_8=0$. Parameters the same as in Fig.~\ref{fig:Quadripartite}a.}
\label{fig:Bipartite}
\end{figure}

It is also interesting to study the behaviour of the bipartite entanglement parameter, $\tilde{S}_{nm}$, between two selected modes among the four being studied. Figure~\ref{fig:Bipartite}a shows the variation of $\tilde{S}_{nm}$, evaluated between pairs of signal modes, for the eight mode ring with all driving fields switched on (the case of Fig.~\ref{fig:Quadripartite}a). It can be seen that there is no bipartite entanglement between any pair of modes since inequality~\ref{eq:S} is not violated - despite the fact that quadripartite entanglement exists, this entanglement is locked in the four entangled modes. This is similar to the property of the Greenberger-Horne-Zeilinger (GHZ) three-qubit state: tracing out one party leaves the system unentangled. Figure~\ref{fig:Bipartite}b shows the variation of $\tilde{S}_{nm}$ for the case when one driving field is removed from the eight mode ring. In this case the system displays neither quadripartite nor bipartite entanglement.

\section{Conclusion}

Solid-state systems of arrays of weakly nonlinear coupled cavities represent a viable route toward the generation of quantum correlated phenomena in a compact device. In this work, we have demonstrated that the emergence of quantum correlations is not necessarily restricted to systems with high cooperativity -- namely large nonlinear energy as compared to the dissipation rate -- but can occur in the weakly nonlinear limit thanks to quantum interference between different pathways of coherent excitation. Using the Wavefunction Monte Carlo approach we were able to study the quantum optical behaviour of moderately large systems. In particular, we have reviewed our early results for the system of three coupled cavities with a driving field acting on the middle one \cite{Liew2012}. This system may be viewed as the elementary building block for the design of more complex arrays for multipartite entanglement generation. We have also extended our previous study by modelling the same array in which the Kerr nonlinearity is replaced by the Jaynes-Cummings one. Also in this case, we have obtained a significant violation of the generalized Bell's inequality, suggesting that this entanglement generation paradigm can be ported to CQED systems such as photonic crystal cavities embedding semiconductor quantum dots \cite{Reinhard2012}. As an example of a system able of generating multipartite entanglement, we proposed an array of eight coupled cavities, which could be realized in a variety of structures including photonic crystals, atom-cavity systems and semiconductor micropillars. We have shown that the entanglement is dependent on the configuration of driving fields and that it can also be generated in systems with other topologies, such as a linear chain of seven modes. This result represents a very important proof of principle for the control of the entanglement, where one would be able to alter the system topology by electric~\cite{Liew2010b} or magnetic~\cite{Zhang2009} fields as well as vary the pattern of the incident optical field. The system has more potential for scalability and control than previous schemes based on parametric scattering~\cite{Liew2011}.

Future work should have, as an objective, to find a deterministic link between the type of entangled state on one side, and the array topology and optimal value of parameters on the other side, similarly to what has been done for the elementary system of two coupled cavities \cite{Bamba2011}. The full numerical solution of the steady-state Von Neumann equation for the density matrix is a very demanding computational task and the arrays studied here are the largest systems that can be tackled by exact numerical methods. The study of larger systems will require the adoption of approximated methods able of correctly describing the quantum correlated nature of the states. Matrix product operators could be the election method for one-dimensional arrays \cite{Hartmann2009,Verstraete2004}, while arrays of higher dimensions would require a generalized tensor network approach \cite{Vidal2008}. Finally, further investigations are still required to assess the usefulness of the entangled states within the field of quantum technology. Indeed, schemes of quantum computation based on multipartite entangled states~\cite{Raussendorf2001} remain promising and have been adapted for continuous variable systems~\cite{Menicucci2006}. However, the criteria for multipartite entangled states to be useful resources have not been completely developed.

\ack

Our work was supported by NCCR Quantum Photonics (NCCR QP), research instrument of the Swiss National Science Foundation (SNSF).

\section*{References}
\bibliographystyle{unsrt}
\bibliography{Refs}

\end{document}